\magnification=1200
\parskip 12pt

\centerline{\bf Book Review}

\centerline{\bf Bohmian Mechanics and Quantum Theory: An Appraisal.} 

\centerline{\bf \ \ \ \ \ \ \ \ \ \ \ James Cushing, Arthur Fine and 
Sheldon Goldstein (Eds.)}

                                 Gregg Jaeger
                                 Boston University,
                                 Dept. of Elect. and Comp. Eng.    
                                 Photonics Center,
                                 8 St. Mary's St,
                                 Boston, MA 02215

     This collection ``Bohmian Mechanics and Quantum Theory'' gives us
the broadest perspective yet on this important realist alternative 
to standard quantum theory. The book treats the history, philosophical 
ramifications and consistency of theories arising from Bohm's original model 
and their relations to other alternative theories. Significantly, it also 
contains applications of these theories to practical situations, such as 
scattering theory - a sign of the field's increasing maturity. 

Most importantly among its contributions to physics and the philosophy
of physics, Bohmian mechanics provides an existence proof in support of
the hidden-variable  approach to explaining quantum phenomena: it is a 
causal mechanics observationally equivalent to standard nonrelativistic 
quantum theory. As with the study of the foundations of quantum theory 
generally, Bohmian mechanics is undergoing a renaissance. During those 
long years of neglect between the appearance of the EPR argument in 1935 
and the emergence of various theories of microphysics, Bohmiam mechanics 
and Everett's relative state theory emerged as little-regarded testimonies 
that alternatives to standard quantum theory were indeed conceivable. In 
these later times, we have this exciting volume of contemporary work on 
Bohmian mechanics. Though the editors have included among the contributors 
several contemporary advocates of the approach, sufficient sceptical analysis 
is also provided to put optimistic claims in proper perspective.

The simplest way to view the introduction in 1952 of Bohm's theory is as a 
direct response to the acausality that Bohr and Pauli, amongst others, saw 
as essential to quantum phenomena. After all, these workers saw acausality 
unmistakably manifest in the apparent impossibility of understanding quantum 
phenomena with theories attributing definite trajectories in space-time. 
Bohm's move was to take as fundamental the contrary, that physical 
systems, beginning with single particles, have definite positions at all 
times and that their initial positions at a given time determine their future 
behavior, in accordance with Newton's equation of motion. This required 
mainly the addition of a ``quantum potential'' to those potentials
ordinarily present. The result was a complete theory of motion for 
microscopic systems whose statistics accord with the predictions of standard 
quantum mechanics. In a broader sense, Bohm's theory can be seen as a 
materialist response to the mystic character of the standard theory, most 
clearly exhibited in its dependence on the special status of the observer 
of quantum phenomena.

The paper by Millard Baublitz and Abner Shimony 
gives order to the conceptual tug-of-war now surrounding Bohmian 
mechanics, distinguishing two components within Bohm's initial theory:
the ``causal view'' and the ``guidance view.'' These contributors show how 
the presence of the two views within Bohmian mechanics engenders an
inner tension that has haunted the subject since its inception. The former, 
causal view involves a Newtonian interpretation of Bohm's equation of 
motion, which includes the auxiliary ``quantum potential.''  The guidance view 
instead makes the ``guidance condition'' -- Bohm's special assumption
that a particle's velocity can be written as the derivative of its 
Hamilton-Jacobi function divided by its mass -- fundamental and exact, 
while dispensing with the quantum potential. Support for the stripped down, 
guidance view is strong, numbering amongst its advocates John S. Bell. By 
contrast, there are not any advocates of the causal view as a stand-alone 
framework. Rather, the causal view is a component of the specific hidden 
variables theories proposed by David Bohm and B.J. Hiley and others.

Under the causal view,  Bohmian mechanics is subsumed under classical 
mechanics -- provided the  additional ``quantum potential'' is included 
-- thereby inheriting the realist metaphysics of classical theory. The 
price paid is the loss of necessary agreement with the Born rule and of
explanations for the behavior of entangled systems. This, in turn, means 
the loss of explanations for measurement results. Such a loss is reason
enough to abandon the formalism of the causal view as sufficient for a
stand-alone theory. The guidance view, on the other hand, reproduces 
the quantum mechanical predictions for entangled systems. The price paid 
is the loss of the intuitive clarity of a classical theory. These two views 
complement one another; abandoning either involves a significant loss, 
one physical and the other intuitive. It is, therefore, understandable 
that Bohm was compelled to incorporate both views into his theory to the 
extent allowed, though this threatens the theory's consistency. This 
tension occupied Bohm, and has occupied others ever since the theory's 
introduction.

Not surprisingly, this situation has also not been resolved within the 
current volume. However, the book does bring us closer to the heart of 
the matter through its careful analyses of existing theory and 
presentations of exploratory variants. Beyond these questions, others 
of more explicitly historical and philosophical natures are also 
addressed in this collection. Recall that another concern of EPR's 
critique of standard quantum theory, beyond the issue of 
completeness, was the nature of ``physical reality.'' Mara Beller's 
essay, ``Bohm and the `Inevitability' of Acausality'' takes up this 
question. Beller carefully delves into Bohm's position, which sees 
a positivistic eschewing of unobservables as the source of 
physicists' reluctance to go beyond the trappings of the standard theory 
since its formulation. (Notably, critic Robin Collins takes a similar 
position against Bohmian realism in his ``Epistemological critique...''.) 
Importantly, Beller points out a long antagonism between Bohm and Bohr 
regarding the relation between classical and quantum concepts: Bohm 
had resisted from the beginning of his career the conclusion that the 
quantum world is fundamentally acausal. It was only later that Bohm's 
theory demonstrated the deniability of the acausality advocated by Bohr.

As Arthur Fine notes in his essay, Bohm developed an independent ideological 
framework based on ``radical holism.'' Fine also points out that the realism 
of Bohm is of an unusual sort; the properties found in measurement are 
not simply disclosed by the act of measuring but may change during 
measurement. This is was recognized by Einstein already in 1953, when he 
saw that in the case of a particle traveling between two walls the 
pre-measurement speed would be zero, with a non-zero velocity being acquired 
during measurement. Such compromised realism turned Einstein away 
from support of the Bohm model.

The question of how the properties of a Bohmian mechanical system are to
be attributed is taken up by Harvey Brown, Andrew Elby and Robert 
Weingard. Ontologically, two components are generally associated with 
a Bohmian system: the $\Psi$-function and the corpuscle. The former has been 
viewed by Peter Holland as acting on the latter in a causal manner. Holland 
assumes the nonlocalizability hypothesis: that the dynamical 
state-independent parameters of mass, charge and magnetic moment cannot 
be attributed to (i.e. localized within) the corpuscle alone. This
hypothesis is motivated in part by the (nonlocal) $\Psi$'s dependence on these
parameters. One who accepts the nonlocalizability hypothesis faces the 
options of parsimony or generosity with regard to $\Psi$'s bearing these 
attributes. On the former option all these properties are associated with 
$\Psi$, whereas on the latter they are attributed to the conjunction of $\Psi$ 
and the corpuscle. Brown et al. suggest a choice in favor of generosity, 
as parsimony would give rise to paradoxical situations such as a pair of 
particles being associated with a single corpuscle should the pair have 
coincident trajectories. 

For his part, Holland argues elsewhere quantum mechanics and Bohmian 
mechanics cannot be universal physical theories since the formal structure 
of quantum theory prevents the recovery of the full range of possible 
classical mechanical motions. In essence, Holland is denying the 
reducibility of Classical to Quantum Theory. In a related contribution,
Detlef Duerr, Sheldon Goldstein and Nino Zanghi argue that Bohmian 
mechanics, being a hidden-variables theory, should be viewed
as the "foundation of quantum mechanics" in that one may arrive at a
version of Bohmian mechanics by adding to quantum mechanics particle 
positions as hidden variables. This raises the question as to whether 
Bohmian mechanics and quantum mechanics or Bohmian mechanics and classical 
mechanics are to be seen as closer pairings.

Variant verions of Bohmian mechanics are also presented, along with factors
motivating them. Trevor Samols points out that, though Bohmian mechanics can 
be viewed as a realist version of quantum mechanics, there is great difficulty
in extending the theory in a relativistic form, however well the phenomenology
(in the physicists' sense) may work out: Bohmian mechanics bears the burden of
a preferred frame of reference due to its use of the guidance condition. 
Samols offers an alternative field theory in the spirit of Bohmian mechanics 
that does not depend on such a frame. Similarly, Antony Valentini offers his 
own "pilot-wave theory" of fields, gravitation and cosmology. P.N. Kaloyerou, 
and Chris Dewdney and George Horton also discuss the treatment of bosonic 
fields within the Bohmian tradition.

In addition to the above contributions, this collection contains a number
of carefully worked out practical applications of Bohmian mechanics
to interferometry, position measurements, scattering theory, tunneling 
phenomena and other practical situations to which any mature physical
theory must be applicable. Though this book does not answer all 
the questions one is compelled to ask of Bohmian mechanics, it does  
demonstrate that Bohm's theory and its successors form a vibrant part
of contemporary physical theory and a locus of stimulating philosophical 
inquiry.

\end